\documentclass[11pt]{article}
\usepackage{graphicx,cite}
\renewcommand{\baselinestretch}{1.0}
\renewcommand{\appendix}
        {
        \par
        \setcounter{section}{0}
        \setcounter{subsection}{0}
        \gdef\afterthesectionpunctdefault{:}
        \gdef\thesection{{Appendix \Alph{section}}}
        \renewcommand{\theequation}{\Alph{section}\arabic{equation}}
        \setcounter{equation}{0}
        }
\begin{document}
\begin{center}
{\huge Rough droplet model for}\break
{\huge spherical metal clusters}\break

\vspace{1.0cm}

{\Large Nicolas Pavloff and Charles Schmit}
\end{center}

\vspace{0.5 cm}

\noindent Division de Physique Th\'eorique\footnote { Unit\'e de
Recherche des Universit\'es Paris XI et Paris VI associ\'ee au CNRS.},
Institut de Physique Nucl\'eaire, F-91406 Orsay Cedex, France \break

\vspace{0.5 cm}
\begin{center}
{\bf abstract}
\end{center}

	We study the thermally activated oscillations, or capillary waves,
of a neutral metal
cluster within the liquid drop model. These deformations correspond to a
surface roughness which we characterize by a single parameter $\Delta$. We
derive a simple analytic approximate
expression determining $\Delta$ as a function of
temperature and cluster size. We then estimate the induced effects on shell
structure by means of a periodic orbit analysis and compare with recent data
for shell energy of sodium clusters in the size range $50 < N < 250$. A small
surface roughness $\Delta\simeq 0.6$ \AA~ is seen to give a reasonable
account of the decrease of amplitude of the shell structure observed in
experiment. Moreover -- contrary to usual Jahn-Teller type of deformations --
roughness correctly reproduces the shape of the shell energy
in the domain of sizes considered in experiment.

\vspace{5cm}
\noindent PACS numbers~:\hfill\break
\noindent 36.40.+d Atomic and molecular clusters.\hfill\break
\noindent 71.20.--b Electron density of states.\hfill\break
\noindent 03.65.Sq Semiclassical theories and applications.\hfill\break

\noindent IPNO/TH 97-19 \hspace{0.5cm} {\it to appear in Phys.
Rev. B}\hfill\break
\newpage

\section{Introduction}

	Since the discovery of shell effects in metal clusters, the
mean field
approach with delocalized electrons has been a very efficient tool for
describing a wide variety of phenomena~: shell and supershell effects, dipole
polarizability and optical excitations, fission \ldots (for reviews see
Refs.~\cite{Bra93,Nah97}).

	This type of approach mainly relies on the jellium model where the
ionic background is considered as a smooth and uniform
distribution of positive
charges. It is best legitimate for simple metals (and to a lesser extend for
noble metals) with delocalized valence electrons, almost insensitive to the
actual arrangement of the ionic cores. Hence the best candidates for this
approximation are the alkali metals, as can be inferred from the
quasi-sphericity of their Fermi surface (revealing a weak interaction between
ionic cores and valence electrons).
As a result the compressibility of these solids
is close to its electron gaz value, and the surface tension is correctly
described by the jellium models (the agreement being better for small
electronic density, see Ref. \cite{Lan73}).

\

	It was realized early that deformation effects had to be taken into
account for a realistic des\-crip\-tion of metal clusters (see {\it eg.} the
review \cite{Deh93}). Within density functional theories, this
can be achieved
by imposing to the jellium the shape that suits the electrons best
\cite{Eka88,Hir94,Kos95}. Deformations have also been studied in
less elaborate
models (deformed external mean-field)
\cite{Cle85,Rei93,Bul93,Rei94,Yan95,Fra96} and there
is a good overall agreement with experimental data for ionization potentials,
dissociation energies and splitting of dipole resonances for relatively small
clusters (less than 40 atoms).

\

	The above mentioned deformations are of Jahn-Teller type and occur
between major shell closures, where lowering the symmetry
leads to a gain in energy.
Another type of surface deformation has also to be considered 
which consists in
surface irregularities of very large multipolarities. These deformations not
only lower the shell effect but also introduce randomness into
the spectrum. This
was first noticed by Gor'kov and Eliashberg \cite{Gor65} who claim that ``the
distribution of the levels should be random even if the particles have
the same
volume and a good shape, say spherical particles of equal size. The point is
that electrons in the metal have a wavelength of the order of atomic
dimensions. Therefore surface irregularities of atomic size are sufficient to
make the level distribution perfectly random". This statement has to be
tempered in view of the success of the jellium model. Nevertheless it
makes no
doubt that the surface of a cluster has atomic size irregularities, it is
important to estimate their amplitude and to evaluate the resulting
effect on the physical observables.

\

	The fact that disorder is located on the surface is
legitimated because the elastic mean free path of an electron in the bulk
is typically of order of several hundreds of Angstr\"oms, whereas an
electron experience collisions
on the surface of the cluster about each 10 \AA. Indeed it was
shown in Ref. \cite{Nis90} that the scattering of electrons on the
fluctuation of the positive ions had an effect of the same order as
that of thermal distribution of occupancy probability~; which in turn
is shown
in the present work to be negligeable compared to the effect of shape
fluctuations. More microscopically, bulk disorder would be represented by
fluctuations of the bottom of the potential well and in the large size limit
high lying states tend to be insensitive to this perturbation, whereas the
effects of surface disorder increase when one goes up in the
spectrum \cite{Pav95}. Note that such
irregularities are to be taken into account also when the cluster is
``liquid-like"~: the mean velocity of the ionic cores is always by several
orders of magnitude smaller than the typical electronic Fermi velocity. 
Hence, as far as electronic motion is concerned, the ionic cores
can be considered as
frozen and this necessarily implies a certain degree of surface roughness.

\

	In the present paper we use a liquid drop model to study the
thermally activated surface deformations, or capillary waves, of a neutral
spherical cluster (Sec. 2). These deformations correspond to a surface
roughness which
we characterize by a single parameter $\Delta$. We derive a simple analytic
approximate
expression determining the behaviour
of $\Delta$ as a function of temperature and cluster size.
Then in Section 3 we discuss the influence of shape fluctuations on the level
density using a trace formula in rough billiards and compare with
thermal effects linked with Fermi occupation number of the energy levels (for
this purpose we give the general form of shell corrections in the presence of
temperature in Appendix C). Finally we present our conclusions and discuss
possible refinements of our approach in Sec. 4.

\section{Liquid drop model}

	In the liquid drop model, a cluster is described as a droplet
of incompressible fluid whose shape can be parameterized by a set of normal
coordinates $\alpha_{\lambda\mu}$ obtained by expanding the surface in
spherical harmonics \cite{Ray45,BM75}~:

\begin{equation}\label{ld1}
r(\Omega,t) = R \left[ 1 + \sum_{\lambda\mu} \alpha_{\lambda\mu}(t) \;
Y_{\lambda\mu}(\Omega) \right] \; .
\end{equation}

	The r.h.s. of Eq. (\ref{ld1}) is made real by imposing
$\alpha_{\lambda,-\mu}=(-)^\mu \alpha^*_{\lambda\mu}$. The
summation stops at a Debye cutoff
$\Lambda$ estimated by equating the number of surface modes to the number of
atoms on the surface, this yields
$\Lambda=(3\sqrt{4\pi}N)^{1/3}\simeq 2.20\,N^{1/3}$.
The droplet being considered as incompressible one should impose
volume conservation. If the cluster contains $N$ atoms it should
have a volume
$V=4\pi R^3/3$ with $R=r_{\! \scriptscriptstyle S}\, N^{1/3}$ 
($r_{\! \scriptscriptstyle S}$ being the Wigner-Seitz radius of
the material). This leads to the relation $\alpha_{00}\sqrt{4\pi} = -
\sum_{\lambda\mu} |\alpha_{\lambda\mu}|^2$, valid to leading order. Also the
modes $\lambda=1$ which correspond to a global translation of the drop should
be omitted in the summation (\ref{ld1}).

\

	Eq. (\ref{ld1}) yields a kinetic energy $T$ and a surface energy
$V_{\mbox{\scriptsize surf}}=\sigma {\cal A}$, where
$\sigma$ is the surface tension and
${\cal A}$ the surface area corresponding to (\ref{ld1}).
Including terms up to
second order in the $\alpha$'s one obtains \cite{Ray45,BM75}~:

\begin{equation}\label{ld2}
T = \frac{\rho_0\, R^5}{2} \sum_{\lambda \mu} \frac{|\dot{\alpha}_{\lambda
\mu}|^2}{\lambda} \qquad\mbox{and}\qquad
V_{\mbox{\scriptsize surf}} = 4\pi R^2\sigma +
\frac{R^2 \sigma }{2}\sum_{\lambda \mu}|{\alpha}_{\lambda \mu}|^2
(\lambda-1)(\lambda+2) \; .
\end{equation}

\noindent where $\rho_0$ is the specific mass of the material considered.

	One can also take into account a curvature term in the
potential energy

\begin{equation}\label{ld3}
V_{\mbox{\scriptsize curv}} = \frac{\gamma}{4}\int dA
\left( \frac{1}{{\cal R}_1} + \frac{1}{{\cal R}_2}\right)
\; .
\end{equation}

	In (\ref{ld3}) $\gamma$ is an intrinsic curvature energy parameter,
${\cal R}_1$ and ${\cal R}_2$ are the principal radii of curvature.
It turns out (see Appendix A) that taking this term into account exactly
amounts to
replacing in Eq. (\ref{ld2}) the surface tension by an effective term
$\sigma\to\sigma^* = \sigma+\gamma/(2 R)$ with which we will work
henceforth.

	No other contribution to the potential energy has to be taken into
account because we consider concomitant deformations of the jellium
and of the
valence electron cloud of a neutral cluster (hence there is no
other electrostatic deformation energy than the one included in
(\ref{ld2},\ref{ld3})). We neglect at this level finite size quantum effects.
We use for the surface tension $\sigma$ the value of the bulk material
extracted from experiment in Ref.~\cite{Tys77}, and this implicitly
contains quantal effects associated with the kinetic energy of the electrons
near the surface. Hence including quantum effects in the present description
would double count this contribution~; the appropriate procedure would be to
use a Strutinsky shell correction, {\it cf.} the discussion at the end of the
paper. In Ref.~\cite{Per91} a description analogous to the present one
has been
shown to account accurately of the monovacancy formation energy in
simple metals such as the one we are interested in. This gives us confidence
in the ability of a liquid drop model to describe atomic size
irregularities. Note that in \cite{Per91} the value of $\sigma$ is
renormalized in order to describe
an ideally flat surface. This procedure should not be employed here
because we
want the surface tension of a large cluster to tend to the one of the bulk
material.

\

	Eqs. (\ref{ld2}) and (\ref{ld3}) correspond to a liquid
drop Lagrangian
${\cal L}_{\scriptscriptstyle LD} [\dot\alpha_{\lambda\mu} \, ,
\alpha_{\lambda\mu}] = T - V_{\mbox{\scriptsize surf}}
- V_{\mbox{\scriptsize curv}}$ with normal modes
$\alpha_{\lambda \mu}(t)=A_{\lambda \mu} \exp (i\omega_\lambda t)$ of
pulsation $\omega_\lambda$ given by~:

\begin{equation}\label{ld4} \omega_\lambda^2 =
\lambda(\lambda-1)(\lambda+2)\frac{\sigma^*}{\rho_0\, R^3} \; ,
\end{equation}

\noindent and the classical energy of the mode is $E_{\lambda\mu} =
\sigma^* R^2(\lambda-1)(\lambda+2)|A_{\lambda \mu}|^2$. The
average value of the amplitude $|A_{\lambda \mu}|^2$ of the thermally
activated mode is determined by writing $E_{\lambda\mu} =
k_{\! \scriptscriptstyle B} T$. 
We use here classical statistical mechanics, the quantal analog would
be $E_{\lambda \mu}= \hbar\omega_\lambda(n_\lambda+1/2)$, where $n_\lambda =
[\exp(\hbar\omega_\lambda/k_{\! \scriptscriptstyle B} T) - 1]^{-1}$ is
a Bose occupation factor. Such a
description has been used for describing the surface oscillations of liquid
helium \cite{Col70}, but here the motion of the surface is
classical~: $k_{\! \scriptscriptstyle B} T\gg \hbar\omega_\Lambda$ 
(from (\ref{ld4})
$\hbar\omega_\Lambda\simeq 130$ K for sodium).

	From (\ref{ld1}) the quantity $r(\Omega,t)$ averaged over the surface
has a mean value $R(1+\alpha_{00}/\sqrt{4\pi})$ and a standard deviation
$\Delta$ which is is given by $\Delta^2 =
R^2\sum_{\lambda\ge 2} |\alpha_{\lambda\mu}|^2/(4\pi)$, hence $\Delta$ is
independent of time. The explicit formula reads~:

\begin{equation}\label{ld6}
\Delta^2 = \frac{\displaystyle k_{\! \scriptscriptstyle B} T}
{\displaystyle 4\pi\sigma^*}
\sum_{\lambda=2}^{\Lambda}\frac{2\lambda+1}{(\lambda-1)(\lambda+2)}
\simeq \frac{\displaystyle k_{\! \scriptscriptstyle B} T}
{\displaystyle 4\pi\sigma^*}
\; \ln \; \frac{(2\Lambda-1)(2\Lambda+5)}{7} \; .
\end{equation}

	In the r.h.s. of Eq. (\ref{ld6}) we replaced the discrete
summation by
the first term of its Euler-MacLaurin expansion. Figure (1a)
displays the result
of Eq.~(\ref{ld6}) for sodium clusters at temperatures
$T=200$ K and $T=450$ K in the
size region $20\le N \le 1000$. It is difficult to determine 
the precise value
of $\sigma^*$ to be used in Eq.~(\ref{ld6})~: the surface and curvature
parameters $\sigma$ and $\gamma$ depend on temperature and of the actual
phase
(liquid or solid) of the aggregate. Hence we used several values
of $\sigma$ and $\gamma$~: a lower bound for $\Delta$ is
obtained by taking the values $\sigma = 190$ K.\AA$^{-2}$ (which
is the solid-vapor value extrapolated to zero
temperature in Ref.~\cite{Tys77}) and $\gamma=285$ K.\AA$^{-1}$ \cite{Per91}.
The upper bound is obtained by taking $\gamma=0$ and 
$\sigma=145$ K.\AA$^{-2}$
(which is the liquid-vapor surface tension at melting \cite{Tys77}).
These values of $\sigma$ correspond to a droplet
parameter $a_s=4\pi r_{\! \scriptscriptstyle S}^2\sigma$
which ranges from $0.68$ eV (for $\sigma=145$ K.\AA$^{-2}$) to $0.89$ eV (for
$\sigma=190$ K.\AA$^{-2}$). Indeed one can find a large dispersion of
$a_s$ in
the literature~: the value 0.54 eV was used in Refs.~\cite{Bra89,Yan95} (from
a fit to theoretical values of clusters' energy)~; in Ref.~\cite{Nah97} the
value $a_s=0.7$ eV was extracted from the bulk surface tension and in Ref.
\cite{Bre94} the value $a_s=1.02$ eV was obtained {\it via} experimental
determination of clusters' cohesive energy.

\begin{figure}[thb]
\begin{center}
\includegraphics*[width=12cm,bbllx=10pt, bblly=80pt, bburx=565pt,
bbury=530pt]{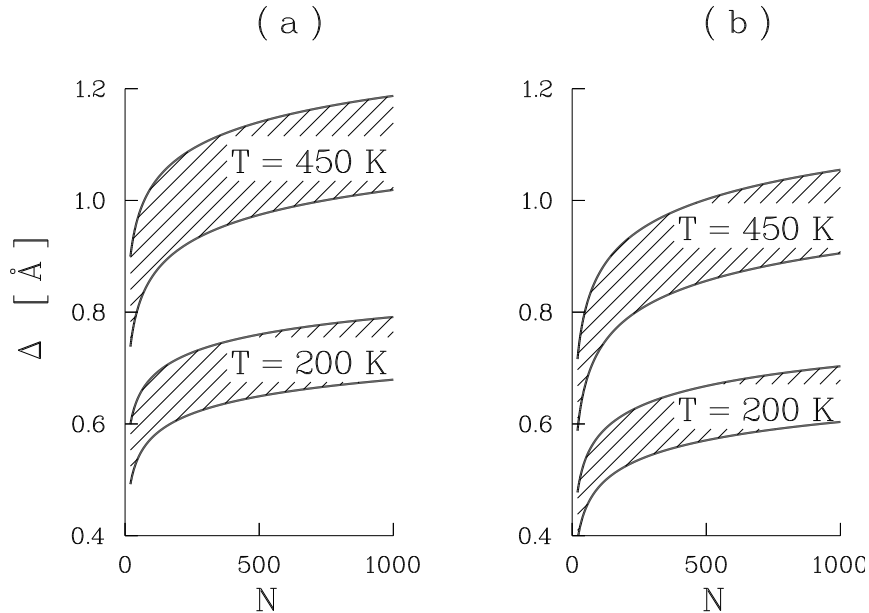}
\end{center}
\caption{Part (a)~: $\Delta$ as a function of $N$ for
sodium clusters at temperatures
200 K and 450 K, from Eq.~(\ref{ld6}). The shaded zones correspond to
different values of the surface tension and curvature parameter (see the
text). Part (b)~: same as part (a) withdrawing in
Eq.~(\ref{ld6}) the contribution of the two first multipolarities
($\lambda=2$ and $\lambda=3$).}
\end{figure}

	As several studies in the field have demonstrated
(see {\it e.g} the work of Yannouleas and Landman \cite{Yan95,Yan96})
the liquid-drop Lagrangian
${\cal L}_{\scriptscriptstyle LD}$ is not adequate for describing the
deformations of small multipolarities, which are determined mainly by shell
effects (see the discussion at the end of the paper). In order
to estimate the role of these multipolarities in the amplitude of the
roughness we have redrawn Figure~(1a) withdrawing the contributions of
$\lambda=2$ and 3 in the summation (\ref{ld6}). The result is displayed on
part (b) of Fig.~1. As one should expect
this significantly reduces the value of $\Delta$.
For instance, at $T=$ 450 K
the typical roughness for $N\simeq 200$ is reduced from 1 \AA~ to 0.85~\AA.

\

	From the discussion above and the comparison of Figures (1a)
and (1b),
we see that due to the simplicity of our model we
cannot precisely determine the
surface roughness of sodium clusters at typical ex\-pe\-ri\-men\-tal
temperatures. For sodium $r_{\! \scriptscriptstyle S}=2.08$~\AA~ (thermal
variations of $r_{\! \scriptscriptstyle S}$ are
negligeable here), and we can only say that typical roughnesses are 
of order of 30 to 50 \% of the interatomic distance. However
it makes no doubt
that there is a thermal activation of capillary waves which gives a
contribution to the surface roughness
of the type given by Eq. (\ref{ld6}) for large enough
multipolarities. We will see below that
this has an important effect on shell structure, and that this
effect is crucial
for understanding the shape of the shell energy determined in experiment.

	Note that an estimation of the
geometrical corrugation of a solid surface at zero temperature \cite{Per91}
yields values of $\Delta$ of order of 20 \% of $r_{\! \scriptscriptstyle S}$.
Hence the reduction of
shell oscillations presented below is a general phenomenon which does not
depend on the solid or liquid structure of the aggregate.

\section{shell energy in a rough sphere}

	The influence of surface roughness on the level statistics has been
discussed in Refs.~\cite{Rat80,Buc90,Pav92,Man93} and more recently in a
2D model \cite{Fra97}. In the present work we concentrate on its
effect on level density and shell structure. Similar effects have been
recently studied in Refs.~\cite{Aku94,Aku95} and \cite{Ler95}. The
spirit of the present section is very similar to the one of 
Ref.~\cite{Ler95} which presents numerical results in a corrugated mean
field. However the qualitative conclusions are different~: in
Ref.~\cite{Ler95} corrugation is seen to imply a
shift in the supershell structure. This effect -- not seen in the present
study -- seems to be due to the fact that, in a finite depth potential
such as the one used in \cite{Ler95}, roughness leads to an effective
mean-field where the phase difference between the orbits is modified.
In any case, for the small roughnesses we are using here, the shift in
the shell structure found in Ref.~\cite{Ler95} is small.
In Ref.~\cite{Aku94} the disorder is modelized by the addition of a random
Hamiltonian to the mean field and the results are comparable to the one
presented below. The approach has been further extended in Ref.~\cite{Aku95}
where the effects of disorder on energetics of lithium, sodium and potassium
clusters where taken into account in a liquid drop plus shell-correction
model. Here the Hamiltonian for the deformation is only of liquid drop
type~; however we do take into account the thermally
activated oscillations for the Hamiltonian we consider (see the discussion in
Sec. 4).

\

	In the present work the $N$ electrons are considered to be moving
independently in an infinite potential well (a billiard) having a shape
approximatively spherical (as given by (\ref{ld1})). Hence we can
consider that
the actual shape is obtained by a random deviation from a perfect sphere,
with
Gaussian fluctuations of standard deviation $\Delta$ determined above. The
choice of Gaussian fluctuations reflects the fact that the distribution of
each $A_{\lambda\mu}$ is Gaussian (according to classical mechanics). Then
invoking the central limit theorem it can reasonably be considered
that the shape
fluctuations are of Gaussian type. We consider an ensemble of clusters 
(such as
one would expect in a molecular beam) and we will present results
averaged over
this ensemble. The radius $R$ of the average sphere scales
with $N$ so that the
mean electronic density is kept constant and equal to its bulk value~: $R =
r_{\! \scriptscriptstyle S}\, N^{1/3}$.

\

	The level density in a rough billiard with small size surface
irregularities was studied in Ref.~\cite{Pav95} and a semiclassical trace
formula averaged over surface disorder was derived. The important feature of
the level density is the gradual disappearance of shell effects with
increasing
energy~: near the Fermi level the electronic wavelength is of order of the
typical size of the surface defects and the induced shift of the eigenlevels
leads after averaging to a structureless level density. The bottom of the
spectrum is not affected because low lying state have a wavelength
much larger
than the surface perturbations (accordingly, the effect on level
statistics is
different at the bottom of the spectrum and near the Fermi energy
\cite{Pav92}).

	It is shown in Ref.~\cite{Pav95} that the oscillatory
part of the electronic energy $E_{\mbox{\scriptsize shell}}$ (the
so-called shell energy) can be
expressed on average as a sum over classical periodic orbits in a perfect
sphere~:

\begin{equation}\label{ld7}
E_{\mbox{\scriptsize shell}}(N,\Delta) \simeq
{\displaystyle \hbar^2\bar{k}_{\! \scriptscriptstyle F}^2
\over\displaystyle 2 m}
\sum_{PO}
{\displaystyle 2 {\cal A}(\bar{k}_{\! \scriptscriptstyle F})\over
\displaystyle
\bar{k}_{\! \scriptscriptstyle F} L^2}
\sin(\bar{k}_{\! \scriptscriptstyle F} L + \nu\pi/2)
\exp\{-2n(\bar{k}_{\! \scriptscriptstyle F}\Delta)^2\cos^2\theta\} \; .
\end{equation}

 	In (\ref{ld7}) $m$ is the electron mass,
$\bar{k}_{\! \scriptscriptstyle F}$ is the smooth part
({\it i.e.} non-oscillatory) of the Fermi wave-vector, which is to a good
approximation equal to the bulk wave-vector $\kappa_{\! \scriptscriptstyle F}
= r_{\! \scriptscriptstyle S}^{-1} (9\pi/4)^{1/3}$.
$E_{\mbox{\scriptsize shell}}$ in (\ref{ld7}) is a quantity averaged over
surface disorder, but the sum is performed over all the periodic orbits (POs)
of a {\it perfect} sphere (see \cite{Pav95}). $L$ is the length of
a PO, ${\cal A}$ is an amplitude slowly depending on
$\bar{k}_{\! \scriptscriptstyle F}$, $\nu$ is a
Maslov index, $n$ is the number of bounces of the PO on the
sphere and $\theta$
is the bouncing angle~; all these quantities depend on the PO considered, see
Appendix B for further details. When $\Delta=0$, (\ref{ld7}) follows from
Balian and Bloch's trace formula for the sphere \cite{BB72}.
Since $\bar{k}_{\! \scriptscriptstyle F}$ is nearly constant, the
main $N$-dependence in (\ref{ld7}) is due to the scaling of the
cluster's size
according to $R=r_{\! \scriptscriptstyle S}\, N^{1/3}$ ($L$ scales like
$R$, ${\cal A}\propto R^{5/2}$ or $R^2$ for some orbits, see Appendix A).

\begin{figure}[thb]
\begin{center}
\includegraphics*[width=9cm,
bbllx=80pt, bblly=125pt, bburx=460pt, bbury=550pt]{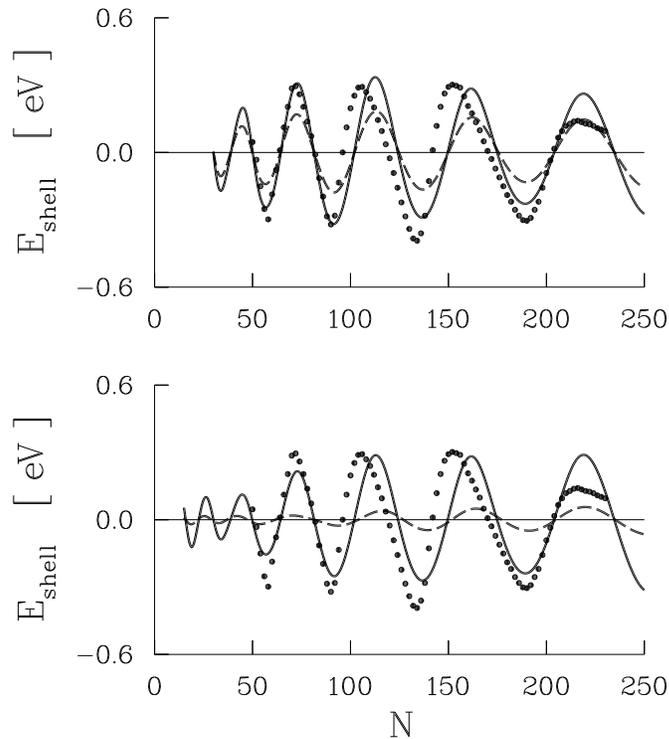}
\end{center}
\caption{$E_{\mbox{\scriptsize shell}} (N,\Delta )$
as a function of $N$ in sodium clusters. The lower
plot compares the experimental results of Ref.~\cite{Cha97}
(black points) with the values obtained by fixing $\Delta=0.9$
\AA~(dashed line) and $\Delta=0.63$ \AA~(solid line).
The upper plot compares the experimental data with the values obtained by
determining for each cluster size $\Delta$ {\it via} (\ref{ld6}), taking
$\sigma=190$ K.\AA$^{-2}$~and $\gamma=285$ K.\AA$^{-1}$,
withdrawing the contributions of $\lambda=2$ and 3. The dashed line
corresponds to a temperature $T=400$ K and the solid line to $T=300$ K.}
\end{figure}

	$E_{\mbox{\scriptsize shell}}$ as given by (\ref{ld7}) is
compared on Fig.~2 with the value
determined by Chandezon {\it et al.} in Ref.~\cite{Cha97}.
In this reference an evaporation model is used for extracting the shell
energy from the abundance
distribution in a beam and results for clusters of sizes ranging
from 50 to 230 atoms are obtained. This experiment is important for our study
because it concerns relatively large clusters and our approach is limited to
this domain for the two following reasons~: (i) we use a semiclassical
approach more accurate for large sizes (see for instance
the comparison with exact results on Fig. 4) and
moreover (ii) the macroscopic concept of
roughness is meaningless for small sizes~; for instance in a cluster with
$N=20$ atoms the Debye cut-off fixes the 
maximum angular momentum of surface deformation to be
$\Lambda \simeq 2.2 \, N^{1/3} \simeq 6$ and in this regime
the concept of roughness is of marginal importance (see below the
discussion of Jahn-Teller effect). In the following we state rather loosely
that our approach is relevant for sizes
$N \hbox{\lower .8ex\hbox{$\, \buildrel > \over \sim\,$}} 100$.

	On the lower part of Fig.~2 the dashed
curve corresponds to a constant value $\Delta=0.9$ \AA~ (independent of $N$)
which -- from Fig.~1 --
is a typical value at $T=450$ K (this temperature is in agreement with usual
evaporation conditions \cite{Ley89,evap1,evap2}). The fact that this value of
$\Delta$ leads to a too large
damping of shell structure should not worry us at
this level~: as stated in the previous section, the liquid drop model
does not accurately determine the value of $\Delta$ because it does not
properly describe deformations of small multipolarities. These
deformations should be described with a more
elaborate procedure and may be less easily
thermally excited (see the discussion of Sec.~4). This is confirmed by the
solid curved which is drawn for $\Delta = 0.63$ \AA~ 
and which gives a better account of the
data. Note the sensitivity of $E_{\mbox{\scriptsize shell}}$
to the value of 
$\Delta/r_{\! \scriptscriptstyle S}$~:
formula (\ref{ld7}) has a schematic large $N$ behaviour of the form~:

\begin{equation}\label{ld7b}
E_{\mbox{\scriptsize shell}}(N,\Delta) \sim
\varepsilon_{\! \scriptscriptstyle F} N^{1/6}
\exp(- \Delta^2/r_{\! \scriptscriptstyle S}^2) \sin (N^{1/3}) \; .
\end{equation}

\noindent where $\varepsilon_{\! \scriptscriptstyle F}
=\hbar^2\kappa_{\! \scriptscriptstyle F}^2/(2m)$
is the bulk Fermi energy. For
clarity we have dropped in the sine and the exponent of (\ref{ld7b})
important
but dimensionless factors. This will be done also in (\ref{ld8b})~; the
derivation of these formulae is explained in Appendix C (Eq. (\ref{b12})).
Hence the shell structure is very sensitive to a small surface
roughness~; this point
and the validity of formulae of type (\ref{ld7}) have been further
tested on a numerical example in Ref.~\cite{Pav98}.

\

	The value $\Delta=0.63$~\AA~ does not quite correspond to the
estimation of Fig.~1 for $T=450$ K~: as explained in the conclusion the
liquid drop Lagrangian seems to underestimate the stiffness of the potential
for the deformation parameters.
It may also happen that the electrons
experience a mean-field which -- due to the diffuseness of its surface -- is
less corrugated than the ionic background. In the same line,
instead of fixing $\Delta$ to a constant value, one should according
to (\ref{ld6}) take size-dependence of the roughness into account.
This improves the agreement with experiment for $N\simeq 50$ since in
this region the value of $\Delta$ decreases significantly (see Fig.~1)
and this leads to
lower damping of the theoretical curve which comes closer to experiment.
This has been done on the upper part of Fig.~2 which is drawn in the
case $\sigma=190$ K.\AA$^{-2}$~and $\gamma=285$ K.\AA$^{-1}$~; these values
have been chosen because they lead to small values of $\Delta$ and to
relatively good agreement with experiment. The dashed line correspond to
$T=400$ K and the solid line to $T=300$ K. For each
temperature and cluster size $\Delta$ was determined {\it via} (\ref{ld6}),
withdrawing the contributions of $\lambda=2$ and 3.
However, such a refinement is unnecessary for larger cluster sizes
in view of the small $N$-dependence of
(\ref{ld6}) for large $N$. Besides, the
simple model with a constant value of $\Delta=0.63$ \AA~ gives already
a satisfactory agreement with experimental data.

\

	In fact our approach is more strongly supported by the very good
comparison of theory an experiment for
the {\it shape} of the curve for the shell energy than for the
comparison with the {\it amplitude}. Indeed the agreement with the amplitude
may not be as good as presented on Fig.~2 because there should be
some room left for an extra
reduction of the amplitude of the shell energy due do quantum mechanically
driven deformations, corresponding to relatively small multipolarities
($\lambda=2$ or 3 typically). Nevertheless, due to the high
sensitivity of formula
(\ref{ld7}) to small changes of $\Delta$, we still can conclude from
Fig.~2 that the typical roughness is of order of 0.6 \AA.

	Concerning the shape of the curve, the experimental results of
Ref.~\cite{Cha97} are surprising because they are in contradiction with 
the common belief that cluster's deformations are only governed by 
Jahn-Teller effects. For instance, the
dissociation energies and ionization potentials of simple metal clusters of
relatively small sizes are well accounted for by models where the
Jahn-Teller effect is the only mechanism of deformation \cite{Yan95,Yan96}
(the agreement with experiment survives up to size $N\sim 100$ for the
ionization potentials of potassium, see Ref.~\cite{Yan97}).
This phenomenon was expected to occur even for large values of
$N$, see {\it e.g.} the zero temperature
results of Refs.~\cite{Rei94,Fra96,Rei93,Bul93}, or the finite temperature
results of Ref.~\cite{Fra96a}. In these studies, the
deformations occur between shell closure and their main effect is to remove
the upper part of the shell oscillations~; the shell energy is predicted
to have sharp negative spikes in vicinity of the magic numbers (these spikes
correspond semiclassically to long POs).
On the other hand surface roughness suppresses long POs and
reduces shell structure more
uniformly, as seen in the experimental data of Ref.~\cite{Cha97}. Hence
we feel that previous theoretical approaches overestimate the role
of the Jahn-Teller mechanism~:
the very specific shape of shell energy they predict
is not seen in the experiment of Chandezon
{\it et al.} The comparison between our approach and
the experimental results for the sizes
$N \hbox{\lower .8ex\hbox{$\, \buildrel > \over \sim\,$}} 100$
firmly establishes that there is a qualitatively
important effect of roughness.

	On the quantitative level, one can also notice that
typical theoretical studies overestimate the shell effect
for clusters of large sizes~: compare Fig. 3 of Ref.~\cite{Cha97} with
similar figures of Refs.~\cite{Rei94,Fra96,Rei93,Bul93}. Temperature
effects improve the agreement (see Ref.~\cite{Fra96a}) but
there is still a mismatch of
order of 40 \% for the amplitude of shell oscillations (see Fig. 3 of
Ref.~\cite{Cha97}), leaving room for improvement due
to surface roughness.

\

	For further comparison with experimental data we
display on Table~1 the magic numbers in the region $N<1300$. The first column
shows the $\Delta=0$ results from the semiclassical formula (\ref{ld7}). The
second column displays the exact result in the perfect sphere
and merely tests
the accuracy
of the semi-classical periodic orbit expansion used in the first column.
Note that the magic numbers of these two columns are almost
identical to the results of Bulgac and Lewenkopf \cite{Bul93} who use a
quadrupole deformation of a spherical billiard model within the shell
correction method~: this is due to the fact that, as stated above, there is
no Jahn-Teller deformation at shell closure. In the third column we show the
magic numbers obtained from (\ref{ld7}) with $\Delta=0.63$ \AA. The three
first
columns compare well with the experimental ones from Chandezon {\it et al.}
(column 4), and roughness has only a small effect on the location of
the magic
numbers. We still produce these data because they justify
the billiard model we are using~: the magic numbers from Table~1 are
in better agreement with experiment than the one obtained with harmonic
oscillators \cite{Rei93,Rei94} or more elaborate potentials
\cite{Fra96,Fra96a}. Hence, as far
as the phase difference between the contribution of its POs to (\ref{ld7}) is
concerned, the billiard model is presumably close to the experimental
situation
since it allows a good prediction of the minima in the shell energy.
However the electrons are sensitive to a mean-field which -- due to the
finite range of the electron/ion and electron/electron interaction -- could
be less corrugated than the ionic background. This would lead
to an effective decrease of $\Delta$ and may help improving the model by
reducing the importance of the ionic corrugation, leaving some
room for an extra decrease in amplitude due to deformations of Jahn-Teller
type. 

\

\begin{table}
\begin{center}
\begin{tabular} {|c|c|c|c|} \hline\hline
 & & & \\
 Eq. (\ref{ld7}), $\Delta=0$ & Exact result, $\Delta=0$ &
 Eq. (\ref{ld7}), $\Delta=0.63$ \AA~ & 
 \hspace{0.5 cm} Ref. \cite{Cha97} \hspace{0.5 cm} \\
 & & & \\ \hline\hline
 & & & \\
 56       & 58       & 56       & 58          \\
 92       & 92       & 92       & 92          \\
 138      & 138      & 136      & 138         \\
 184      & 186      & 190      & $192\pm 2$  \\
 252      & 254      & 252      & $256\pm 2$  \\
 336      & 338      & 334      & $334\pm 2$  \\
 436      & 440      & 430      & $430\pm 2$  \\
 540/554  & 542/556  & 526      & $540\pm 5$  \\
 610/674  & 612/676  & 624      & $648\pm 5$  \\
 744/830  & 748/832  & 752     &   \\
 908      &   912    &  902    &   \\
 1070     &  1074    & 1082    &   \\
 1282     &  1250    & 1286    &   \\
& & & \\ \hline\hline
\end{tabular}
\end{center}
\caption{Magic numbers in the perfectly spherical billiard (column 1~: PO
expansion, column 2~: exact results) and in the rough billiard (column 3).
Column 4 shows the experimental results of Chandezon {\it et al}.}
\end{table}

	Note that in the present treatment the effects of temperature are
indirect~: although the usual
temperatures reached in experiments are small compared to the Fermi
energy (one
remains in the highly degenerate limit
$k_{\! \scriptscriptstyle B} T\ll \varepsilon_{\!\scriptscriptstyle F}$) they
are sufficient to induce a disorder of the cluster's shape
which has a sizeable effect on shell structure. For
comparison one can derive a formula (similar to (\ref{ld7})) encompassing the
effect of a Fermi occupation function in the energy levels of the electron
gas. The free energy $F(N,T)$ is more appropriate than the total energy
for evaluating these effects. Indeed, based on
Weisskopf's approach, the electronic contribution to the evaporation rate of
a neutral monomer from a cluster of size $N$ 
can be shown to be approximatively proportional to
$\exp\{[ F(N)-F(N-1)]/k_{\! \scriptscriptstyle B} T\}$ \cite{Fra96a,Han94}.
The general formula for the oscillating part of the free energy is
derived in 
Appendix C and reads in the case of a billiard~:

\begin{equation}\label{ld8}
F_{\mbox{\scriptsize shell}}(N,T) \simeq
{\displaystyle \hbar^2\bar{k}^2_\mu\over\displaystyle 2 m}
\sum_{PO}{\displaystyle 2 {\cal A}(\bar{k}_\mu)\over
\displaystyle \bar{k}_\mu L^2}
\sin(\bar{k}_\mu L + \nu\pi/2) F_1(\bar{X}) \; ,
\end{equation}

\noindent where $\bar{k}_\mu$ is the non-oscillatory part of the quantity
$k_\mu$ defined by $\mu=\hbar^2k_\mu^2/2m$, $\mu$ being the che\-mi\-cal
po\-ten\-tial. Again $\bar{k}_\mu$ is to a good approximation 
equal to the bulk
Fermi wave vector $\kappa_{\! \scriptscriptstyle F}$.
$\bar{X}=(\pi/2) \tau L \kappa^2_{\! \scriptscriptstyle F}/\bar{k}_\mu$
is a dimensionless quantity which tends to zero at $T=0$
($\tau=k_{\! \scriptscriptstyle B} T/ \varepsilon_{\! \scriptscriptstyle F}$
is the reduce temperature).
More precisely
it can be considered as small if the thermal wave
length $\lambda_T=(2\pi\hbar^2/mk_{\! \scriptscriptstyle B} T)^{1/2}$ is
large compared to
$\sqrt{r_{\! \scriptscriptstyle S} L}$
($\bar{X}=2\pi^2 L/\bar{k}_\mu\lambda_T^2$). $F_1$ is a dimensionless
damping function defined in Appendix C (Eq.~(\ref{c4})).

	We compare on Fig.~3 the effects of a temperature $T =$ 750~K,
with those of a constant roughness
$\Delta= 0.63$ \AA.
The shell energy is displayed as a function of $N^{1/3}$
for sodium clusters of size $N<3400$.
There is a striking difference with the $N$-dependence obtained
{\it via} usual
temperature effects on occupation numbers. As one notices from the figure,
roughness damps the oscillations in the total energy with an overall
factor of the type $\exp (-\Delta^2/r_{\! \scriptscriptstyle S}^2)$
({\it cf} (\ref{ld7b}))
without modifying the qualitative features of the supershells~; whereas
temperature leads to a $N$-dependent damping of schematic form ({\it cf}
Appendix C, Eq.~(\ref{b12}))~:

\begin{equation}\label{ld8b}
F_{\mbox{\scriptsize shell}}(N,T) \sim
\varepsilon_{\! \scriptscriptstyle F}
N^{1/6} F_1(\tau N^{1/3})
\sin(N^{1/3}) \; , \end{equation}

\noindent (as in Eq.~(\ref{ld7b}) we
have omitted numerical factors in the sine and $F_1$ function). Hence
the effects of thermal distribution of occupation numbers is to wash out the
beating pattern of the shell energy by exponentially
damping the large $N$ oscillations (see also Fig.~4).

\begin{figure}[thb]
\begin{center}
\includegraphics*[width=12cm,
bbllx=1pt, bblly=135pt, bburx=560pt, bbury=550pt]{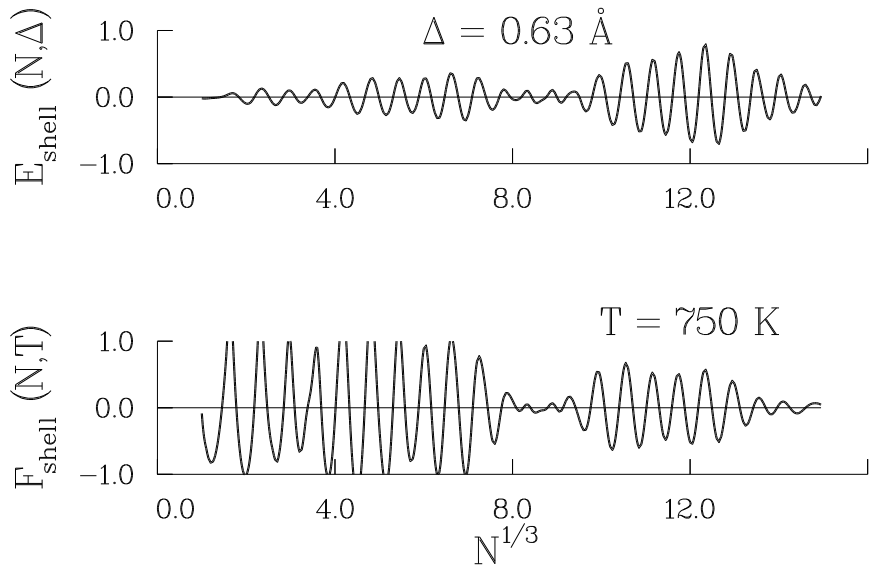}
\end{center}
\caption{$E_{\mbox{\scriptsize shell}}$ (expressed in eV)
as a function of $N^{1/3}$ in sodium clusters.
The upper plot is obtained by taking a  constant
roughness $\Delta= 0.63$ \AA.
The lower plot corresponds to formula (\ref{b12a}) for a
perfectly spherical aggregate with a temperature
$\tau=k_{\! \scriptscriptstyle B} T/\varepsilon_{\! \scriptscriptstyle F}
=0.02$, {\it i.e.} $T=750$ K.}
\end{figure}

	Note also the efficiency of a small roughness for diminishing
shell effect~: for clusters of size $800< N < 1200$, from Eqs.
(\ref{ld7}) and (\ref{ld8}), one can see for instance that the effect of a
small roughness $\Delta = 0.2\; r_{\! \scriptscriptstyle S} \simeq R /50$
on $E_{\mbox{\scriptsize shell}}$ is similar to the one
of a temperature of about 550 K on $F_{\mbox{\scriptsize shell}}$.
In the range $300 < N < 500$ the same
roughness corresponds to a temperature $T\simeq 750$ K, and for
$50 < N < 250$ it corresponds to $T\simeq 1000$ K (around $N=200$) or 1400 K
(around $N=100$). As a result, in a range of sizes and
temperatures commonly reached in experiment ($N\sim 500$ and $T\sim 400$ K)
the effect of roughness on shell structure is dominant compared to that of
thermal distribution of occupation numbers. Furthermore, as discussed above,
it seems from the experimental results of Ref~\cite{Cha97} that in about the
same region, Jahn-Teller deformations of small multipolarity play a smaller
role in the shape of the shell energy than predicted by usual theoretical
studies (see Fig.~3 of Ref.~\cite{Cha97}). We present here roughness as a
concomitant
phenomenon which (according to comparison with the data of Ref. [34])
seems more relevant for large clusters. As discussed in next section both
phenomena should be taken into account for a proper description of
deformations of large clusters.

\section{Discussion}

	One of the original interests of metal clusters was
to provide a physical
realization of a discrete and random spectrum. It was long thought that
the randomness of the levels would lead to a structureless level density
and theoretical works were mostly devoted to the study of the two point form
factor of the spectrum \cite{Per81,Hal86}. It is only during the
last two decades that molecular beams have made it possible
to work in a regime where the size of the cluster is well defined and
small compared to the electronic mean free path. In this regime one
observes shell effects as a result of finite size quantum effects
\cite{Bra93,Deh93}. Nevertheless the remark of Gor'kov and Eliashberg quoted
in the introduction remains valid to some extend and the present work aims to
reconcile these two views by providing a semi-classical description of shell
structure in presence of shape disorder.

\

	The model we have considered is schematic~; a more elaborate
procedure would be to design a Lagrangian for the surface deformation
encompassing the effects of shell structure (in the spirit of 
Strutinsky shell corrections)~:

\begin{equation}\label{last}
{\cal L}  =
{\cal L}_{\scriptscriptstyle LD} [\dot\alpha_{\lambda\mu} \, ,
\alpha_{\lambda\mu}] - E_{\mbox{\scriptsize shell}}
[\alpha_{\lambda\mu}] \; .
\end{equation}

	This approach is typical in the study of deformations of finite
fermionic systems. It is used for determining the equilibrium shape,
{\it i.e.}
the set of $\alpha_{\lambda\mu}$ minimizing the total potential energy.
Between shell closure it
leads to a ground-state in which the equilibrium value
of some of the $\alpha$'s is non-zero (mainly for small $\lambda$),
contrary to
what is obtained in Sec.~2 where all the $\alpha$'s are zero at equilibrium.
Using the Lagrangian (\ref{last}) would also modify the stiffness of the
potential near the minimum (since a term $E_{\mbox{\scriptsize shell}}$ would
be added to the potential used in Sec. 2).

	Following the procedure of the present paper one
should go one step further and study the thermally activated vibrations of
the normal modes in the Hamiltonian (\ref{last}). Hence the usual
Jahn-Teller deformations correspond to the first
step of the procedure just exposed and to small multipolarity whereas
surface roughness to the second step (and to large multipolarity).

	It would be of great interest to verify if the agreement obtained in
Sec.~3 with experimental values would persist when describing surface
oscillations with a Lagrangian such as ${\cal L}$ defined in
Eq.~(\ref{last}). This form of Lagrangian is legitimated by confrontation
with experiments in the small and intermediate size domain, where it is
commonly admitted that clusters of size
$N\hbox{\lower .8ex\hbox{$\, \buildrel < \over \sim\,$}} 100$
experience static deformations of small multipolarities \cite{Yan95,Yan96}.
The success of the
present model (which uses ${\cal L}_{\scriptscriptstyle LD}$ and do not
include such Jahn-Teller deformations) in the size range
$N\hbox{\lower .8ex\hbox{$\, \buildrel > \over \sim\,$}} 100$ might
be explained by the decrease of the shell-energy contribution to ${\cal L}$
due to an intrinsic roughness of the surface. Schematically one
might say that there is less difference between a rough sphere 
and a rough ellipsoid
than between a perfect sphere and a perfect ellipsoid. A similar
phenomenon explains the disappearance of Jahn-Teller deformations
with increasing temperature, see Ref.~\cite{Fra96a} where deformation is
seen to be suppressed by thermal fluctuations. Note however that
the phenomenon predicted in this reference is size-dependant, {\it i.e.}
not uniform for all cluster sizes (as roughness would
be), see the precise discussion in \cite{Fra96a}.

\

	Note finally that thermally induced shape fluctuations have already
been investigated in the study of the broadening of plasma
resonances of metal
clusters in Refs.~\cite{Ber89,Pac89,Yan90,Pen90} and very recently in the
study of shell structure for clusters of size smaller than $N=100$
\cite{Yan97}. The ideas are similar to the
one exposed above, however the allowed deformations are limited to simple
shapes, whereas it has been considered necessary in the present work to
include deformations of very large multipolarities for studying surface
roughness.

\

\noindent {\large \bf Acknowledgments}

\noindent We thank an unknown referee for pointing out the existence of
Ref.~\cite{Din52}. We thank S. Frauendorf, J. Lerm\'e 
and W. Swiatecki for fruitful
discussions. We wish also to express our gratitude to S. Bj{\o}rnholm for
his interest in this study, judicious remarks and inspiring comments. La
Division de Physique Th\'eorique de l'Institut de Physique Nucl\'eaire est
une unit\'e de recherche des Universit\'es Paris XI et Paris VI associ\'ee au
CNRS.

\appendix

\section{}
\setcounter{equation}{0}

      In this appendix we compute the curvature energy 
$V_{\mbox{\scriptsize curv}}$ (defined in Eq. (\ref{ld3})) for a droplet of
shape given by (\ref{ld1}). This amounts to evaluate the integral
${\cal C} =\int dA(1/{\cal R}_1 + 1/{\cal R}_2)$ for a boundary
approximatively spherical. This can be done by noticing
that if ${\cal A}$ is the surface
area of a given boundary, the modification $\delta{\cal A}$ caused by an
infinitesimal displacement of the boundary reads
$\delta{\cal A}=\int dA\,\delta\zeta (1/{\cal R}_1 + 1/{\cal R}_2)$, where
$\delta\zeta$ is the normal segment between the undeformed boundary and the
deformed one (see {\it eg.} \cite{LL}, chap. VII). If this
modification corresponds to
a modification of Eq. (\ref{ld1}) by $r\to r+\delta r$, one can compute
$\delta{\cal A}$ and $\delta\zeta$ in terms of $\delta r$. This allows to
write ${\cal C}$ in the form

\begin{equation}\label{gdif1}
{\cal C} = \int d\Omega \; \frac{K}{r}
\left\{ \frac{r^2+K}{K^{1/2}} -
\frac{1}{\sin\theta}\, \partial_\theta
\left(\frac{\displaystyle r\sin\theta\, \partial_\theta r}
{\displaystyle K^{1/2}}\right)
-\frac{1}{\displaystyle \sin^2\theta}\, \partial_\phi
\left(\frac{\displaystyle r\, \partial_\phi r}
{\displaystyle K^{1/2}}\right)
\right\}
\; , \end{equation}

\noindent where $K(\Omega)=r^2+(\partial_\theta r)^2
+(\partial_\phi r)^2/\sin^2\theta$. Then, writing $r(\Omega)=R\, [
1+h(\Omega)]$ and neglecting terms of order greater than ${\cal O}(h^2)$ one
obtains :

\begin{equation}\label{gdif2}
{\cal C} = 2 R \int d\Omega \,\left\{
1 + h +\frac{1}{2}(\partial_\theta h)^2 +
\frac{\displaystyle 1}{\displaystyle 2\sin^2\theta}\,
(\partial_\phi h)^2\right\}
+ {\cal O}(h^3) \; . \end{equation}

	The surface area can be expressed in a similar
manner (see \cite{LL}) :

\begin{equation}\label{gdif3}
{\cal A} =  \int dA = \int d\Omega\; r \, K^{1/2}
= R^2 \int d\Omega \,\left\{
(1 + h)^2 +\frac{1}{2}(\partial_\theta h)^2 +
\frac{\displaystyle 1}{\displaystyle 2\sin^2\theta}\, 
(\partial_\phi h)^2\right\}
+ {\cal O}(h^3) \; .
\end{equation}

	The condition of volume conservation
imposes $\int\! d\Omega\, (1+h)^2=\int\! d\Omega\, (1+h) +{\cal O}(h^3)$.
Hence, comparing Eqs. (\ref{gdif2}) and (\ref{gdif3}) one sees that
for small deformations the curvature integral is proportional to the
surface area~: ${\cal C} = 2 {\cal A}/R + {\cal O}(h^3)$. The corrections are
of third order in the deformation, they are given in Ref. \cite{Has88}
(chap. 6) for spheroidal and harmonic deformations. Here we are interested
only in terms up to order ${\cal O}(h^2)$, thus the curvature energy
$V_{\mbox{\scriptsize curv}}=\gamma\, {\cal C}/4$ is equal up to a
multiplicative constant to the surface term
$V_{\mbox{\scriptsize surf}}=\sigma {\cal A}$~: $V_{\mbox{\scriptsize curv}}$
is formally obtained from $V_{\mbox{\scriptsize surf}}$ by replacing
the surface tension $\sigma$ by an effective term
$\gamma/(2 R)$.

\section{}
\setcounter{equation}{0}

      In this appendix we briefly present the results of Ref.~\cite{BB72} for
the level density in the sphere and we make explicit the different terms
appearing in Eq. (\ref{ld7}) for the perfect billiard.

	The periodic orbits in the sphere are regular polygons in diametral
planes. They are labeled by two numbers $(n,t)$, $n$ being the number
of sides
and $t$ the winding number of the orbit around the center ($n \geq 2t$). Note
that $n$ is here the same as the number of bouncing points appearing in the
main text (Eq. (\ref{ld7})). The oscillating part of the level density in the
sphere reads~:

\begin{equation}\label{a3}
\rho_{\mbox{\scriptsize osc}}(k) = \sum_{t=1}^{+\infty}
\sum_{n=2t}^{+\infty}
{\cal A}_{n,t}(k) \sin (k L_{n,t} + \nu_{n,t}\pi/2) \; .
\end{equation}

	The shortest orbits are the pendulating orbit ($n=2$, $t=1$), the
triangle
($n=3$, $t=1$), and the square ($n=4$, $t=1$). The triangle and the
square are
sufficient to understand the qualitative features of the shell and supershell
structure (see {\it e.g.} \cite{Ped91,Sta92}). Each orbit bounces on the
surface with a constant normal angle $\theta_{n,t} = (1-2t/n)\pi/2$ and has a
length $L_{n,t}=2 n R \cos\theta_{n,t}$. The pendulating orbit occurs in a
two-parameters family (the parameters determine the direction of bouncing)
whereas all the other orbits form three-parameters families. Hence
the bouncing
ball (with $n=2t$) has to be treated separately. The explicit formulae for
${\cal A}(k)$ and $\nu$ in Eqs.~(\ref{ld7}) and (\ref{a3}) are ($t \ge 1$)~:

\begin{equation}
\label{a1} \nu_{n,t} = \left\{ \begin{array}{ccc}
                                 0 & \quad \mbox{if}\quad & n=2t \; , \\
                             n+3/2 &  \quad \mbox{if}\quad & n>2t \; ,
                                 \end{array}\right.  \end{equation}

\noindent and

\begin{equation}\label{a2}
{\cal A}_{n,t} (k) = \left\{ \begin{array}{lcc}
-{\displaystyle d_{\scriptscriptstyle S}
k R^2\over\displaystyle \pi t}  & \quad \mbox{if}\quad & n=2t \; , \\
2 d_{\scriptscriptstyle S}
(-1)^t \sin (2\theta_{n,t})
\sqrt{{\displaystyle\cos\theta_{n,t}\over\displaystyle\pi n}}
R(R k)^{3/2} &  \quad \mbox{if}\quad & n>2t \; ,
                                 \end{array}\right. \end{equation}

\noindent where $d_{\scriptscriptstyle S}=2$ is the spin degeneracy.

	One sees that the amplitude corresponding to the pendulating orbit is
proportional to $k$ whereas the other families have a larger weight
(proportional to $k^{3/2}$). Generally speaking, one can show
that the contribution of a $d$-parameter family has an extra $k^{d/2}$ power
with respect to that of an isolated orbit \cite{Cre91}.

\section{}
\setcounter{equation}{0}

	In this appendix we derive approximate analytical expressions for the
oscillatory part of the total energy and of the free energy at finite
temperature. Similar results concerning the entropy, the free energy
{\it etc}\ldots have been previously obtained by Kolomiets, Magner and
Strutinsky in Refs.~\cite{Str76,Kol79}. We nevertheless briefly outline the
derivation of the formulae because the references just quoted are not
very explicit and difficult to follow. The formulae are derived in the
framework of a general PO expansion~: the level density is noted
$\rho(\epsilon)$, it is separated in a smooth term $\bar\rho(\epsilon)$
and an oscillating term $\rho_{\mbox{\scriptsize osc}}(\epsilon)$. In the
present work, we denote all the smooth terms with an upper bar and the
oscillating terms with a subscript ``osc'',
except for the oscillating part of the energies which have a subscript
``shell" according to the general convention in the field.
$\rho_{\mbox{\scriptsize osc}}(\epsilon)$ is supposed to be of the form~:

\begin{equation}\label{c1}
\rho_{\mbox{\scriptsize osc}}(\epsilon) =
\, \mbox{Re}\, \sum_{PO} {\cal B}(\epsilon) \,
{\mbox{\large e}}^{\displaystyle iS(\epsilon)/\hbar} \; ,
\end{equation}

\noindent where $S(\epsilon)$ is the action of the PO considered.
${\cal B}(\epsilon)$ is an orbit dependent amplitude which is
of order $\hbar^{-1}$ for chaotic systems, of order
$\hbar^{-2}$ for typical integrable systems in three dimensions and of
order $\hbar^{-5/2}$ for rotationally symmetric systems as
the spherical billiard where
families of orbits are characterized by 3 parameters \cite{Cre91}.

\

	We will estimate the asymptotic form of several integrals, all of the
same type, and we first display a formula often used below. Let
$g(\epsilon)$ be
a slowly varying function of $\epsilon$ (as ${\cal B}(\epsilon)$ is supposed
to be), and $g^\prime$ its first derivative. Let
$\phi(\epsilon-\mu)=[1+\exp\{ (\epsilon-\mu)/
k_{\! \scriptscriptstyle B} T\} ]^{-1}$
be the Fermi function, $\mu$ being the chemical potential.
If $S(\mu )\gg \hbar$ one has~:

\begin{equation}\label{c3}
\int_0^{+\infty}\!\!\!\!
g(\epsilon) \, \phi(\epsilon-\mu) \, 
{\mbox{\large e}}^{i S(\epsilon)/\hbar} d\epsilon
=
\frac{\displaystyle \hbar}{\displaystyle i} \, 
\frac{{\mbox{\large e}}^{iS(\mu)/\hbar}}{\displaystyle S'(\mu)}
\left\{
g(\mu)F_1(X) - \frac{\displaystyle \hbar}{\displaystyle i}
\, \frac{g'(\mu)}{S'(\mu)} F_2(X)
+ \frac{\displaystyle \hbar}{\displaystyle i}
\frac{g(\mu) S''(\mu)}{[S'(\mu)]^2} F_3(X)
+ \cdots \right\} \; , \end{equation}

\noindent where the integral has been evaluated by a contour integration
in the
complex plane (see {\it e.g.} \cite{Ric96}). In the evaluation of the
integral
we have neglected the contribution of a part of the contour located on the
positive imaginary axis, this is legitimate provided the temperature is small
compared to the Fermi energy (degenerate Fermi gas approximation). $X=\pi
S'(\mu)k_{\! \scriptscriptstyle B} T/\hbar$ is a dimensionless
quantity which can be considered as small if the period $S'$ of the orbit
is small compared to a characteristic thermal time
$\hbar/k_{\! \scriptscriptstyle B} T$. $F_1$, $F_2$ and $F_3$ are
dimensionless damping functions~:

\begin{equation}\label{c4}
F_1(X) = {\displaystyle X\over\displaystyle\sinh X} \; ,\qquad
F_2(X) = {\displaystyle X^2\cosh X\over\displaystyle\sinh^2 X} \; ,\qquad
F_3(X) = {\displaystyle X^3\over\displaystyle\sinh^3 X}
(1+{\sinh^2 X\over 2}) \; .
\end{equation}

	For obtaining the total energy starting from (\ref{c1}), one
first determines the chemical potential $\mu$ through the equality
$N={\cal N} (\mu)$, where $N$ is the number of electrons and
${\cal N}(\mu)=\int \rho(\epsilon)\phi(\epsilon-\mu)d\epsilon$. As 
$\rho(\epsilon)$, ${\cal N}$
can be separated in a smooth term $\bar{\cal N}$ (the Weyl term) plus an
oscillating part ${\cal N}_{\mbox{\scriptsize osc}}$. Accordingly, $\mu$ can
be separated in a
smooth function of $N$ plus an oscillating term~:
$\mu=\bar{\mu} + \mu_{\mbox{\scriptsize osc}}$, where
$N=\bar{\cal N}(\bar{\mu})$.

      	Then the total electronic energy is $E=\int \epsilon
\phi(\epsilon-\mu) \rho(\epsilon)d\epsilon$. It can also be separated into a
smooth part $\bar{E}$ and an oscillating part which is denoted
$E_{\mbox{\scriptsize shell}}$ throughout the paper
(more precisely $E_{\mbox{\scriptsize shell}}(N,T)$ in presence of
temperature) in accordance with the
general notations in the field. $E_{\mbox{\scriptsize shell}}$ reads
appro\-xi\-matively~:

\begin{equation}\label{b6}
\begin{array}{ccl}
E_{\mbox{\scriptsize shell}} & = &
\displaystyle\int\epsilon\, \rho(\epsilon)\phi(\epsilon-\mu)d\epsilon  -
\int\epsilon\, \bar\rho(\epsilon)\phi(\epsilon-\bar\mu)d\epsilon   \\
& \simeq & \displaystyle\int\epsilon\, \rho(\epsilon)\,
[\phi(\epsilon-\bar\mu)-\mu_{\mbox{\scriptsize osc}}
\, \phi'(\epsilon-\bar\mu)]d\epsilon -
\int\epsilon\, \bar\rho(\epsilon)\phi(\epsilon-\bar\mu) d\epsilon \\
& = &
\displaystyle\int \epsilon\, \rho_{\mbox{\scriptsize osc}}
(\epsilon) \phi(\epsilon - \bar\mu)d\epsilon
+\mu_{\mbox{\scriptsize osc}}
\, \bar\mu\left(\frac{\displaystyle d{\cal N}}{d\mu}\right)_{\bar\mu}
-\mu_{\mbox{\scriptsize osc}}
\int(\epsilon-\bar\mu)\rho(\epsilon)\phi'(\epsilon-\bar\mu)d\epsilon
\; .
\end{array}
\end{equation}

	The last term of the r.h.s. of (\ref{b6}) is subdominant, moreover
it is zero at zero temperature, hence we drop it in the following.
Then, from (\ref{c1}) and (\ref{c3}) one obtains~:

\begin{equation}\label{zobi}
E_{\mbox{\scriptsize shell}}(N,T) \simeq - \,\mbox{Re}\, \sum_{PO}
\left(\frac{\displaystyle\hbar}{\displaystyle iS'(\bar\mu)}\right)^2
{\cal B}(\bar\mu)F_2(\bar{X})\,
{\mbox{\large e}}^{\displaystyle iS(\bar{\mu})/\hbar} \; ,
\end{equation}

\noindent where $\bar{X}$ is computed as $X$ with $\bar\mu$ replacing
$\mu$.

\

	The free energy $F(N,T)$ is a quantity more appropriate than the
total energy to evaluate the effects of electronic tem\-pe\-ra\-tu\-re
on the
abundance of clusters in the beam (see the discussion in the main text).
It is defined by

\begin{equation}\label{fe1}
F(N,T) = \mu N + \int_0^{+\infty}\!\!\!
d\epsilon \, \rho(\epsilon) \Phi(\epsilon-\mu) \; \,
\qquad\mbox{where}\qquad
\Phi(\epsilon-\mu) = - k_{\! \scriptscriptstyle B} T
\ln\left(1+
{\mbox{\large e}}^{\displaystyle(\mu-\epsilon)/k_{\! \scriptscriptstyle B} T}
\right) \; .
\end{equation}

	The oscillating part of the free energy is denoted by
$F_{\mbox{\scriptsize shell}}(N,T)$ and can be evaluated similarly to what
has been done in (\ref{b6}). This yields~:

\begin{equation}\label{fe4}
F_{\mbox{\scriptsize shell}}(N,T) \simeq - \,\mbox{Re}\, \sum_{PO}
\left(\frac{\displaystyle\hbar}{\displaystyle iS'(\bar\mu)}\right)^2
{\cal B}(\bar\mu)F_1(\bar{X})\,
{\mbox{\large e}}^{\displaystyle iS(\bar{\mu})/\hbar} \; ,
\end{equation}

	In the particular case of a billiard whose level density is
of the type (\ref{a3}), Eq. (\ref{fe4}) reads~:

\begin{equation}\label{b12a}
F_{\mbox{\scriptsize shell}}
\simeq {\displaystyle \hbar^2\bar{k}^2_\mu\over\displaystyle 2 m}
\sum_{PO}{\displaystyle 2 {\cal A}
(\bar{k}_\mu)\over\displaystyle \bar{k}_\mu L^2}
\sin(\bar{k}_\mu L + \nu\pi/2) F_1(\bar{X}) \; ,\end{equation}

\noindent where $k_\mu$ is defined by $\mu=\hbar^2k_\mu^2/(2m)$. A formula
of this type seems to have been derived first by Dingle in Ref.
\cite{Din52}. We have
verified that this formula is of very good
accuracy in the spherical billiard (see Fig.~4). An equally
good agreement is
obtained for the comparison of $E_{\mbox{\scriptsize shell}}$ (as given by
(\ref{zobi})) with the exact result. For relatively low values of $N^{1/3}$
(say $N^{1/3}<6$), an even better agreement can
be obtained by still using the semi-classical level density, but evaluating
integrals such as (\ref{fe1}) numerically.

\begin{figure}[thb]
\begin{center}
\includegraphics*[width=12cm,bbllx=1pt, bblly=55pt, bburx=560pt,
bbury=520pt]{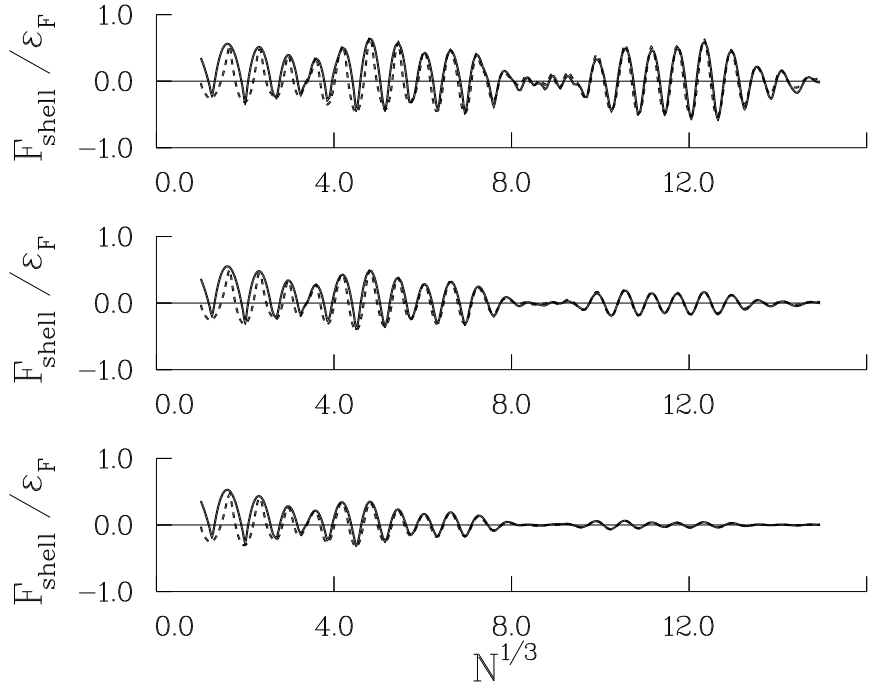}
\end{center}
\caption{$F_{\mbox{\scriptsize shell}}$ (expressed in units of
the bulk Fermi energy $\varepsilon_{\! \scriptscriptstyle F}$) as a function
of $N^{1/3}$ in a perfectly spherical aggregate.
$F_{\mbox{\scriptsize shell}}$ is denoted
$F_{\mbox{\scriptsize shell}}(N,T)$ in the main text. The
different plots correspond to temperatures $\tau=0.01$ (upper plot),
$\tau=0.02$ and 0.03 (lower plots), where
$\tau=k_{\! \scriptscriptstyle B} T/\varepsilon_{\! \scriptscriptstyle F}$.
For sodium these values correspond to $T=376$ K, 752 K and 1128 K.
The solid lines correspond to the determination of
$F_{\mbox{\scriptsize shell}}$ obtained by using the
{\it exact} spectrum of the spherical billiard,
and the dashed lines to Eq.~(\ref{b12a}).}
\end{figure}

	In the sphere, the main
contribution to (\ref{b12a}) comes from orbits occurring in three-parameters
families ({\it cf} Appendix B). Considering that $\bar{k}_\mu$ is of order
$\kappa_{\! \scriptscriptstyle F}\sim 1/r_{\! \scriptscriptstyle S}$ and
that $L$ and $R$ scale
like $r_{\! \scriptscriptstyle S}\, N^{1/3}$ one obtains
the following leading order~: ${\cal A}(\bar{k}_\mu) \sim
R(R\bar{k}_\mu)^{3/2}\sim r_{\! \scriptscriptstyle S}\, N^{5/6}$.
Hence the schematic large $N$ behaviour of (\ref{b12a}) reads

\begin{equation}\label{b12}
F_{\mbox{\scriptsize shell}} \sim
\varepsilon_{\! \scriptscriptstyle F} N^{1/6}
F_1(\tau N^{1/3}) \sin(N^{1/3}) \; . \end{equation}

\noindent where
$\tau= k_{\! \scriptscriptstyle B}T/\varepsilon_{\! \scriptscriptstyle F}$
is the
temperature expressed in units of the bulk Fermi energy. Here we have dropped
for clarity important but dimensionless factors in the sine and $F_1$
function~: we just want to illustrate the typical $N$ dependence
of $F_{\mbox{\scriptsize shell}}$.
We have adopted the same type of notation in the text
(\ref{ld7b},\ref{ld8b}). The behaviour (\ref{b12}) is in agreement with the
findings of  Kolomiets, Magner
and Strutinsky in Refs. \cite{Str76,Kol79} and
with Ref.~\cite{BM75} where Bohr and Mottelson used schematic forms of
the level density. We emphasize that we have used here a generic
PO expansion,
and Eqs. (\ref{zobi},\ref{fe4}) are valid for any system (chaotic
or integrable) of
independent fermions moving in an external potential.

\

	Note finally that we have here computed the thermodynamical
quantities in the grand canonical ensemble. The number of electrons in a
cluster being exactly conserved, the canonical description should be used
instead (hence we should have noted $F(\mu,T)$ instead of $F(N,T)$~: in all
the appendix $N$ should be understood as the mean number of electrons).
The difference between the two ensembles has been
studied in Ref.~\cite{Bra91} where it is shown to give discrepancies of
order of 0.05 eV (or 0.1 eV at best) in the free energy difference
$F(N-1)-F(N)$. This difference
is expected to decrease in the large $N$ limit and moreover
it plays no role in the discussion of the effects of
temperature given in the main text.


\renewcommand{\baselinestretch}{.7}

\end{document}